\documentclass[11pt,twoside,usegraphicx]{article}

\usepackage{asp2010}

\resetcounters

\begin{document}

\title{On time variability and other complications in studying 
the UV broad absorption lines of quasars: 
results from numerical simulations of radiation driven disk winds.}
\author{Daniel Proga$^{1,2}$, Paola Rodriguez-Hidalgo$^{3}$, 
and Fred Hamann$^{4}$
\affil{$^1$Princeton University Observatory, Peyton Hall, Princeton, NJ 08540, USA}}
\affil{$^2$Permanent Address: University of Nevada, Las Vegas, Department of Physics and Astronomy, 4505 S. Maryland Pkwy, Las Vegas, NV 89154-4002, USA}
\affil{$^3$ Department of Astronomy and Astrophysics, Pennsylvania State University, University Park, PA 16802}
\affil{$^4$ Department of Astronomy, University of Florida, Gainesvill, FL 32611}

\begin{abstract}
We review the main results from axisymmetric, time-dependent
hydrodynamical simulations of radiation driven disk winds in AGN. We 
illustrate the capability of such simulations to provide useful
insights into the three domains of observational astronomy: 
spectroscopy, time-variability, and imaging. Specifically,
the synthetic line profiles predicted by the simulations
resemble the broad absorption lines observed in quasars.
The intrinsically time dependent nature of radiation driven disk winds
that have been predicted by the simulations can be supported
by a growing number of the observed dramatic variability in 
the UV absorption lines. 
And finally, the intensity maps predicted by the simulations
give physical and geometrical justification to the phenomenologically
deduced fact that a proper interpretation of the observed line absorption
requires the wind covering factor to be considered as being
partial, inhomogeneous, and velocity dependent.

\end{abstract}

%%%%%%%%%%%%%%%%%%%%%%%%%%%%%%%%%%%%%

\section{Introduction}

AGN are powerful sources of both electromagnetic radiation and mass 
outflows. Broad absorption lines (BALs) in quasars are the most dramatic
evidence of the mass outflows.

Although quasars were discovered 50 years ago, only now are we
starting to gain observational insight into the BAL time-variability
on various time scales by monitoring relatively small samples
of objects \citep[e.g.,][]{Barlow:1994, Lundgren:2007, Gibson:2008, 
Gibson:2010, Capellupo:2011}. The variability of AGN outflows 
can be very dramatic. In particular, it could be manifested as emergence of 
BALs in UV spectra of quasars 
\citep[e.g.,][]{Hamann:2008, Krongold:2010, Hall:2011}
and even in a Seyfert 1 galaxy \citet{Leighly:2009}.
The time-variability provides new constraints for the outflow physics and
geometry. The prospects of studying the time variability of BALs
are very good considering the success of recent observational campaigns
(see Brandt and Haggard in this volume).
For example, spectra of 150,000 quasars from
the Baryon Oscillation Spectroscopic Survey (BOSS) of SDSS-III are expected
to increase the number of known quasars with multi-year variability in BALs
by two orders of magnitude, offering a sample of up to 2000 objects
instead of 20.

Disk accretion onto a massive black hole (BH) is most likely
a source of the powerful radiation. Therefore, it is very plausible that 
the BAL outflows originate from the accretion disks 
and are driven by the same powerful radiation  
\citep[e.g., reviews by][]{Konigl:2006}. 
We argue that disk winds are a crucial ingredient of disks and 
can help us understand the entire disk accreting system.
In particular, if broad emission  lines, one of the defining
features of quasars, are associated with disk winds 
\citep[e.g.,][]{Richards:2011}, then a physical
model of the latter will be a very important element of understanding
reverberation-mapping measurements used to estimate BH masses, 
one of the fundamental parameters of AGN. 
In addition, as mass outflows propagate outward, they can 
significantly affect the medium they interact with. Therefore, 
the outflows  should be  one of the key processes in the so-called AGN 
feedback (see, Ostriker and other contributions in this volume).

%%%%%%%%%%%%%%%%%%%%%%%%%%%%%%%%%%%%

\section{Simulations}

Disk winds are inferred to exist in many, if not even all disk
accreting systems. Therefore, there is an extensive literature
on disk wind studies where various approaches -- such as analytic
and semi-analytical models and numerical simulations -- were
adopted. The problem is quite complex due to non-spherical geometry and 
the richness of the physical processes operating in the disk and wind. 
Therefore, the level of completeness and self-consistency 
varies from model to model.

\begin{figure}
\centering
\includegraphics[width=1. \linewidth]{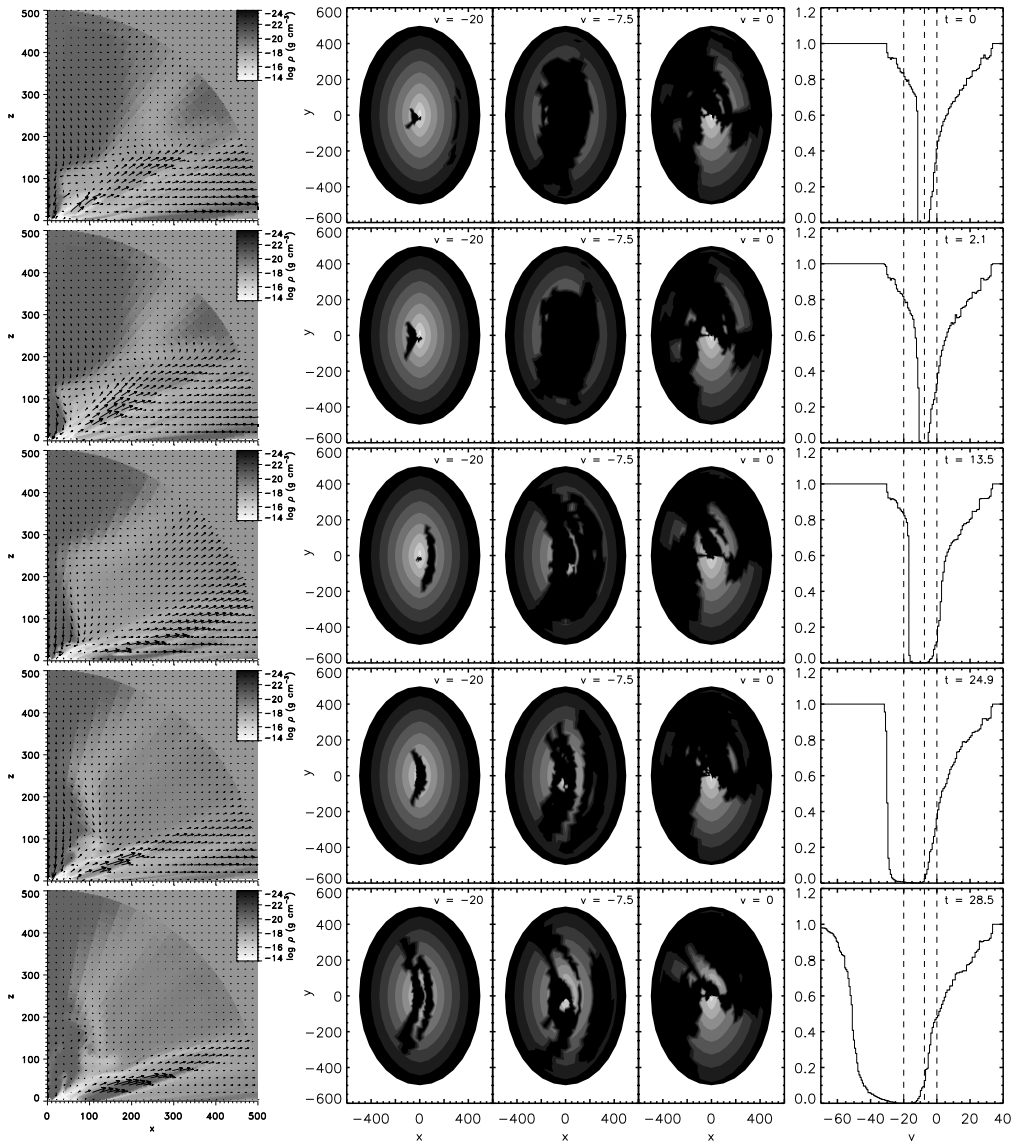}
\caption{A time sequence of density and velocity maps (left column), 
intensity maps (second, third, and fourth columns) and of absorption line
profiles (right column). The time 
(the labels in the top right corner of the right panels)
is measured in months with respect to the time of the first snapshot that 
has been described in detail in \citet{Proga:2004}.
In the density and velocity maps, the rotation axis of the disk is along 
the z-axis, while the midplane of the disk is along the x-axis.
The position on the maps is expressed in units of the disk inner radius
$r_\ast$ which corresponds to 3 Schwarzschild radii (the inner
radius of the computational domain is 10 $r_\ast$). 
The profiles and maps of the disk intensity as viewed through a disk wind 
are computed for the C~IV~1549~\AA\ line and an inclination angle 
of 75$^\circ$ (the angle is measured from the rotational axis). 
Hence the disk in the intensity maps is flattened and
the near and far side of the disk are, respectively
for positive and negative values of the x-coordinate.
The maps are plotted using a logarithmic scale with ten levels
each spanning 2 orders of magnitude. 
The three columns with intensity maps correspond
to three velocities (see the labels in the top right corner of the maps). 
The three velocities are  also marked by 
the dashed vertical lines in the panels with the line profiles.
All velocities in the figure are in units of $10^4~{\rm km~s^{-1}}$.
[The disk wind simulations are from \citet{Proga:2004} while the intensity maps
and line profiles from Rodriguez-Hidalgo et al. (in preparation)].
}
\label{fig:maps}
\end{figure}

We have been involved in several numerical studies of radiation driven 
disk winds. In particular, in \citet{Proga:2000} and \citet{Proga:2004},
we presented results from axisymmetric, time-dependent simulations
of an accretion disk wind driven by radiation pressure on spectral lines.
The simulation, described in greatest detail in \citet{Proga:2004},
is for a nonrotating BH with a mass of 10$^8~{\rm M_{\odot}}$ 
and an accretion luminosity of 50\% of the Eddington luminosity.
It followed three flow components: (i) a hot and low density inflowing gas 
in the polar region, (ii) a dense, warm and fast {\it equatorial}
outflow from the disk, and (iii) a transitional zone in which the disk outflow
is hot and struggles to escape the system. The third component
shields the disk wind (the second component) from powerful ionizing
radiation produced by the central engine so that radiation pressure
on spectral lines can launch and accelerate the wind.
One of the predictions of the radiation-driven disk wind simulations 
is that the wind is intrinsically time-dependent:
the wind solution is unsteady even though the base of the wind
and the driving radiation are assumed to be time independent.
The left panels of Figure~\ref{fig:maps} illustrate the overall wind
structure and time variability.

The simulations of radiation-driven disk winds succeeded in producing 
a fast wind that is capable
of accounting for BAL winds and other winds. Thus they
can be considered  a ``proof-of-concept'' for a radiation driven 
disk wind in AGN. 

To gain more insights into how these simulations compare with
observations,
synthetic line profiles and broad band spectra have been calculated
based on the simulations.
The synthetic line profiles show a strong dependence on inclination
angle: the absorption forms only when
an observer looks at the source through the fast wind (i.e.,
$i \geq 60^\circ$; see fig. 2 in \citet{Proga:2009} ).
This $i$-dependence could explain why only 15\% of quasars have BALs:
it may simply be a selection effect of viewing angle.
The model also predicts high column densities and subsequently
strong X-ray absorption features for the
inclination angles at which strong absorption lines form 
\citep[e.g.,][]{Schurch:2009, Sim:2010}. This trend is
consistent with the observational finding that BAL quasars are
under-luminous in X-rays compared to their non-BAL counterparts
\citep[e.g.,][]{Brandt:2000, Gallagher:2007, Giustini:2008}.

However, these spectral features should vary in time because
they are computed based on the time-dependent wind solutions.
To illustrate this point, \citet{Sim:2010} presented results
from Monte Carlo simulations of the wind photoionization structure
and of the spectra for two snap shots. In the right column of 
Figure~\ref{fig:maps}, we show more examples of this behavior, 
namely synthetic line profiles of a representative UV resonance line, 
C~IV~1549~\AA\, for a fixed inclination angle of $75^\circ$ but at
various times.

The synthetic line profiles indeed change with time but not necessarily
in a fashion one would expect based on just following the variability
of the density and velocity distributions. In particular, the wind structure
is very dynamical at small radii from where the wind is launched and 
the failed wind that shields the outer wind develops (the third flow
component in the simulations that we mentioned above). One could expect
that this inner part of the flow will produce structured absorption
changing on relatively short time scales, of order of days and weeks. 
However, the predicted line profiles
show that the absorption at small velocities is relatively stable
and strong. The most dramatic evolution in the line profiles
occurs at very large velocities that correspond to the emergence
of very fast mass ejections from relatively large distances, where
the gas is well shielded from the X-ray radiation.

To better understand how and where the line absorption takes place,
the second, third, and fourth columns of Figure~\ref{fig:maps} present
maps of disk intensity 
corrected for absorption by C~IV~1549~\AA\ in the wind
[see eq. 3 in \citet{Proga:2002}, albeit with a zeroed source function,
as here we do not model emission].
The maps were computed using the same data that was used to compute
the line profile. In particular, each disk annulus is assumed to radiate
as a black body, with the temperature determined by the BH mass,
the accretion rate, and the annulus distance from the BH.
This disk intensity is then modified by the effects of transmission
through the wind: for a given ray originating a given place on the disk, 
we used the wind solution to look for the resonance points in the wind
where the C~IV~1549~\AA\ absorption occurs at a specific velocity
corresponding to a specific wavelength of continuum emission from the disk
[see \citet{Proga:2002} for more details]. We picked three 
representative velocities: 0, 7.5, and 20 $\times 10^3~{\rm km~s^{-1}}$.

The intensity maps illustrate several important effects that are of
general relevance for the interpretation of observed line profiles.
In particular, the wind covering factor is partial, spatially inhomogeneous, 
and velocity and time dependent. These effects
introduce complications in inferring the wind optical depth,
density and other properties based on observations.
The significance of the effects has been realized by some who
model observed spectra \citep[e.g.,][]{Barlow:1997, Arav:1999, deKool:2002,
Arav:2008}. Thus, our results support the idea that one should
use allow for partial, inhomogeneous, velocity dependent wind covering 
factors while fitting or modelling observed line profiles.
Another related effect is that the intensity maps show no symmetry
even though the disk and its wind are axisymmetric. This effect is due 
to projection and the wind kinematics, i.e., its rotation and expansion.
%%%%%%%%%%%%%%%%%%%%%%%%%%%%%%%%%%%

\section{Concluding Remarks}

The time-dependent, axisymmetric simulations of radiation driven disk
winds have shown that the wind geometry, structure and dynamics are quite
complex. In particular, the wind covering factor is partial, inhomogeneous, 
and velocity and time dependent. Here, we illustrated that the wind 
complexity predicted by the simulations can be and should be accounted for 
while modeling and interpreting the observed absorption lines.

\acknowledgments 

DP acknowledges the UNLV sabbatical assistance and
Support for Program number HST-AR-12150.01-A that
was provided by NASA through
a grant from the Space Telescope Science Institute, which is operated by the
Association of Universities for Research in Astronomy, Incorporated, under
NASA contract NAS5-26555.

\end{document}